\documentclass[11pt,a4paper]{article}
\usepackage[T1]{fontenc}
\usepackage[utf8]{inputenc}
\usepackage{lmodern}
\usepackage[margin=2.3cm]{geometry}
\usepackage{microtype}
\usepackage{amsmath,amssymb}
\usepackage{graphicx}
\usepackage{booktabs}
\usepackage[numbers,sort&compress]{natbib}
\usepackage[hidelinks]{hyperref}
\usepackage{placeins}
\usepackage{float}
\begin{document}
\begin{center}
{\Large\bfseries Depth of Maximum of EAS and Mean Mass Composition of Primary CRs
from the Spatial and Temporal Distribution of Cherenkov Light
According to TAIGA-HiSCORE Data\par}
\vspace{1.1em}
{\footnotesize\setlength{\baselineskip}{11pt}
M.~Yu.~Ternovoy\textsuperscript{1,*},
V.~V.~Prosin\textsuperscript{1}, E.~A.~Osipova\textsuperscript{1},
I.~I.~Astapov\textsuperscript{4}, P.~A.~Bezyazykov\textsuperscript{3},
A.~V.~Blinov\textsuperscript{3}, E.~A.~Bonvech\textsuperscript{1},
A.~N.~Borodin\textsuperscript{3,10}, N.~M.~Budnev\textsuperscript{2},
A.~V.~Bulan\textsuperscript{1}, P.~V.~Busygin\textsuperscript{2},
D.~V.~Chernov\textsuperscript{1}, A.~Chiavassa\textsuperscript{11},
A.~N.~Dyachok\textsuperscript{2}, A.~R.~Gafarov\textsuperscript{2},
A.~Yu.~Garmash\textsuperscript{5,8},
V.~M.~Grebenyuk\textsuperscript{3,10}, E.~O.~Gress\textsuperscript{2},
O.~A.~Gress\textsuperscript{2}, T.~I.~Gress\textsuperscript{2},
A.~A.~Grinyuk\textsuperscript{3}, O.~G.~Grishin\textsuperscript{2},
M.~A.~Iliushin\textsuperscript{2}, A.~D.~Ivanova\textsuperscript{3},
A.~L.~Ivanova\textsuperscript{2}, I.~A.~Kabannik\textsuperscript{5,8},
N.~N.~Kalmykov\textsuperscript{1}, Kh.~Karatash\textsuperscript{3},
V.~V.~Kindin\textsuperscript{4}, S.~N.~Kiryukhin\textsuperscript{2},
K.~G.~Kompaniets\textsuperscript{4},
E.~E.~Korosteleva\textsuperscript{1}, V.~A.~Kozhin\textsuperscript{1},
E.~A.~Kravchenko\textsuperscript{5,8}, A.~P.~Kryukov\textsuperscript{1},
L.~A.~Kuzmichev\textsuperscript{1}, A.~A.~Lagutin\textsuperscript{9},
Yu.~E.~Lemeshev\textsuperscript{2},
B.~K.~Lubsandorzhiev\textsuperscript{6},
N.~B.~Lubsandorzhiev\textsuperscript{1},
S.~B.~Lubsandorzhiev\textsuperscript{6},
A.~D.~Lukanov\textsuperscript{6}, S.~D.~Malakhov\textsuperscript{2},
R.~R.~Mirgazov\textsuperscript{2}, R.~D.~Monkhoev\textsuperscript{2},
A.~L.~Pakhorukov\textsuperscript{2}, A.~T.~Panferova\textsuperscript{1},
L.~V.~Pankov\textsuperscript{2}, A.~A.~Petrukhin\textsuperscript{4},
D.~A.~Podgrudkov\textsuperscript{1}, V.~A.~Poleschuk\textsuperscript{2},
E.~G.~Popova\textsuperscript{1}, E.~B.~Postnikov\textsuperscript{1},
A.~A.~Pushnin\textsuperscript{2}, R.~I.~Raikin\textsuperscript{9},
A.~Yu.~Razumov\textsuperscript{1}, G.~I.~Rubtsov\textsuperscript{6},
E.~V~Ryabov\textsuperscript{2}, V.~S.~Samoliga\textsuperscript{2},
A.~V.~Shaykovsky\textsuperscript{3}, M.~V.~Shulga\textsuperscript{2},
A.~Yu.~Sidorenkov\textsuperscript{6}, A.~A.~Silaev\textsuperscript{1},
A.~A.~Silaev~(Jr.)\textsuperscript{1},
A.~V.~Skurikhin\textsuperscript{1}, A.~V.~Sokolov\textsuperscript{5,8},
L.~G.~Sveshnikova\textsuperscript{1},
V.~A.~Tabolenko\textsuperscript{2}, A.~B.~Tanaev\textsuperscript{2},
P.~A.~Volchugov\textsuperscript{1}, N.~V.~Volkov\textsuperscript{9},
D.~M.~Voronin\textsuperscript{6}, I.~I.~Yashin\textsuperscript{4},
A.~V.~Zagorodnikov\textsuperscript{2}, D.~P.~Zhurov\textsuperscript{2},
V.~N.~Zirakashvili\textsuperscript{7}\par}
\vspace{0.8em}
{\scriptsize
\textsuperscript{1} Skobeltsyn Institute of Nuclear Physics, M.V. Lomonosov Moscow State University, Moscow, 119234 Russia\par
\textsuperscript{2} Research Institute of Applied Physics, Irkutsk State University, Irkutsk, 664003 Russia\par
\textsuperscript{3} Joint Institute for Nuclear Research, Dubna, Moscow oblast, 141980 Russia\par
\textsuperscript{4} National Research Nuclear University MEPhI (Moscow Engineering Physics Institute), Moscow, 115409 Russia\par
\textsuperscript{5} Novosibirsk State University, Novosibirsk, 630090 Russia\par
\textsuperscript{6} Institute for Nuclear Research, Russian Academy of Sciences, Moscow, 117312 Russia\par
\textsuperscript{7} Pushkov Institute of Terrestrial Magnetism, Ionosphere, and Radio Wave Propagation, Russian Academy of Sciences, Moscow, 108840 Russia\par
\textsuperscript{8} Budker Institute of Nuclear Physics, Siberian Branch, Russian Academy of Sciences, Novosibirsk, 630090 Russia\par
\textsuperscript{9} Altai State University, Barnaul, 656049 Russia\par
\textsuperscript{10} Dubna State University, Dubna, 141980 Russia\par
\textsuperscript{11} National Institute of Nuclear Physics, Turin, 10124 Italy\par
\par}
\vspace{0.5em}
{\footnotesize $^*$Corresponding author: \href{mailto:markternovoy96@gmail.com}{markternovoy96@gmail.com}; \\ORCID: \href{https://orcid.org/0009-0006-4826-7080}{0009-0006-4826-7080}\par}
\end{center}
\vspace{0.6em}
\begin{abstract}
We present the dependence of the depth of maximum of extensive air showers
(EAS), $X_{\max}$, on the energy of the primary cosmic particle, as determined
from data recorded with the wide-angle Cherenkov array TAIGA-HiSCORE during
the 2021--2025 observation seasons. The reconstruction of $X_{\max}$ from
the experimental data uses relations obtained from EAS simulations: the
dependence on the steepness of the Cherenkov light lateral distribution
function ($P$-method) and on the full width at half maximum of the
Cherenkov light pulse ($\tau$-method). The reconstructed $X_{\max}$ values
are converted into the mean logarithmic mass $\langle\ln A\rangle$, which
characterizes the primary cosmic-ray mass composition. The results are
compared with measurements from Tunka-133 (2017), the Pierre Auger
Observatory (2019), LHAASO (2024), TALE (2026), and other experiments.

\end{abstract}

\section{Introduction}\label{sec:intro}

This paper describes two methods for determining the depth of maximum,
$X_{\max}$, of extensive air showers (EAS) from data of the
TAIGA-HiSCORE Cherenkov array. The first, the $\tau$-method (or timing
method), is based on the duration of Cherenkov light pulses, measured as
their full width at half maximum (FWHM, $\tau_{1/2}$). The second,
the $P$-method, uses the steepness of the Cherenkov light lateral
distribution function (LDF).

\section{The TAIGA-HiSCORE array}\label{sec:array}

The TAIGA-HiSCORE array is part of the TAIGA Astrophysical complex,
located in the Tunka Valley (Buryatia, Russia) at an altitude of 675~m
above sea level \citep{budnev2022}. It comprises 120 optical stations
arranged on a square grid with a spacing of approximately 100~m,
covering an area of about 1~km$^2$. Each station is equipped with four
photomultiplier tubes (PMTs) and Winston cones, providing a total light
collection area of 0.5~m$^2$ \citep{gress2017}. 
Signals from both the anode and dynode outputs of
the PMTs are recorded to extend the dynamic range
of Cherenkov light measurements.
The signals are digitized
at intervals of 0.5~ns. During processing, signals from the different
PMTs within a station can be shifted by one or more sampling intervals
before summation, allowing a more accurate pulse shape to be recovered.
For each recorded EAS event, the reconstruction determines the shower-core
coordinates in the observation plane, the arrival direction, and the primary-particle energy
\citep{prosin2025,budnev2020}.

The array has operated in its full configuration since 2021. The stations
were pointed toward the zenith during the 2022--2023 and 2024--2025
observation seasons and tilted $25^\circ$ southward during the
2021--2022 and 2023--2024 seasons.

\section{Simulations}\label{sec:sim}

To refine the relation between the observables described below and the
distance to the shower maximum,
$\Delta X_{\max}=X_0/\cos\theta-X_{\max}$, where $X_0$ is the atmospheric
depth at the level of the Tunka Valley ($675~\mathrm{g\,cm^{-2}}$),
EAS simulations were performed using CORSIKA \citep{corsika} with the
QGSJET-II-04 \citep{ostapchenko2011} and Sibyll~2.3d
\citep{riehn2020} hadronic interaction models. Approximately 4,000 EAS
events initiated by primary protons, helium, nitrogen, and iron nuclei
were generated at five energies (1, 3, 10, 30, and 100~PeV) and two
zenith angles ($0^\circ$ and $30^\circ$).

Cherenkov light pulses were simulated with a bunch size of one photon
and without statistical thinning. This increased the computational cost
but improved the accuracy of the simulated pulse shapes. To account for
pulse distortions introduced by the detector electronics (PMTs and
preamplifiers), each simulated pulse was convolved with the instrumental
response function. The mean FWHM of the
instrumental response across all stations is 9.2~ns for the anode
channel and 6.1~ns for the dynode channel.

\subsection{The $\tau$-method}\label{sec:tau}

The timing method (or $\tau$-method \citep{prosin2014tunka}) relies on the calculated dependence
of the width of an EAS Cherenkov light pulse on the relative position of
the shower maximum, $\Delta X_{\max}$. The relationship between
$\Delta X_{\max}$ and the pulse duration $\tau_{1/2}$ at a fixed distance
from the shower axis was obtained by fitting the simulations for four
primary particle types at energies of 10--100~PeV and zenith angles up
to $30^\circ$.

\begin{figure}[H]
\centering
\includegraphics[width=0.78\linewidth]{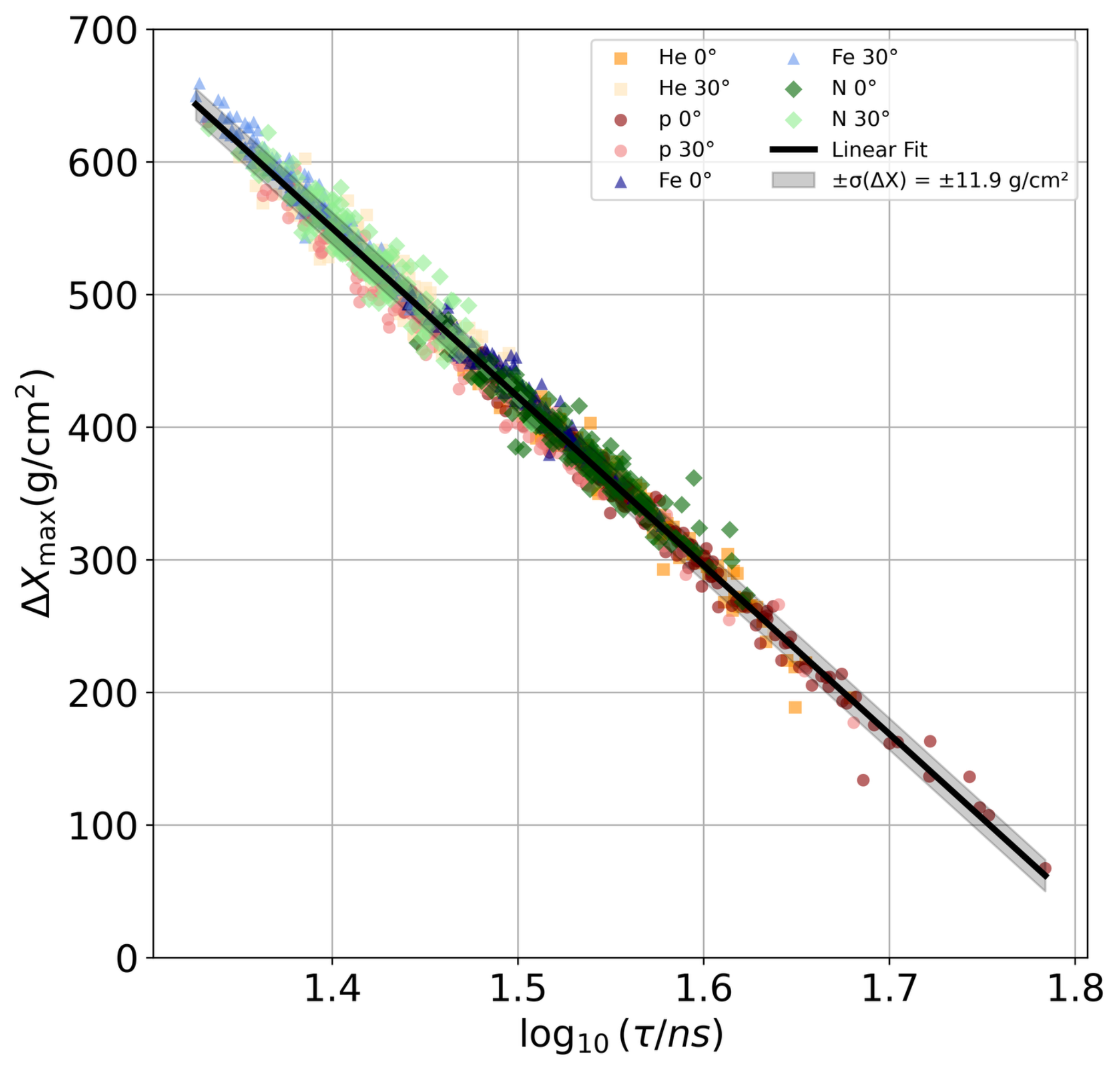}
\caption{Calculated dependence of $\Delta X_{\max}$ on
$\log_{10}(\tau_{300}/\mathrm{ns})$ for the anode channel. The line
shows the linear fit and the shaded band shows its uncertainty.}
\label{fig:tau}
\end{figure}

For a distance of 300~m from the shower axis, fitting the simulated
points yields the following relationship between $\Delta X_{\max}$ and
$\lg(\tau_{300})$ for the anode channel, illustrated in
Fig.~\ref{fig:tau}:

\begin{equation}
\Delta X_{\max}=2330-1272\,\lg(\tau_{300}),\qquad
\sigma=12\;\mathrm{g\,cm^{-2}}.
\label{eq:tau_anode}
\end{equation}

For the dynode channel at the same distance, the corresponding
relationship is

\begin{equation}
\Delta X_{\max}=2009-1085\,\lg(\tau_{300}),\qquad
\sigma=12\;\mathrm{g\,cm^{-2}}.
\label{eq:tau_dynode}
\end{equation}

To determine the mean logarithm of the pulse duration at 300~m experimentally, stations lying 270--330~m from the axis are selected. At all simulated energies, the
pulse durations at these distances exceed the instrumental response
width by at least a factor of two. Moreover, the dependence of the pulse
duration on distance is approximately linear, with nearly the same
slope at all simulated energies. The experimental data provide an
average of 5--10 measurements in the selected distance interval.

\subsection{The $P$-method}\label{sec:P}

The $P$-method \citep{prosin2021}, previously introduced to reconstruct
$\Delta X_{\max}$ from the steepness of the Cherenkov light LDF, has
been modified using the simulation results presented in this work. The
dimensionless parameter is now defined as $P=Q(80)/Q(160)$, providing
a more accurate relation to $\Delta X_{\max}$. Here, $Q(r)$ denotes the
Cherenkov light flux density at a distance $r$ from the EAS axis. The
values $Q(80)$ and $Q(160)$ at distances of 80 and 160~m, respectively,
are obtained by interpolation of the Cherenkov light LDF fitted to the
light densities measured by the detector stations.

The simulated relationship between $P$ and $\Delta X_{\max}$ is

\begin{equation}
\Delta X_{\max}=955-166\,P,\qquad
\sigma=15\;\mathrm{g\,cm^{-2}}.
\label{eq:pcalibration}
\end{equation}

\begin{figure}[H]
\centering
\includegraphics[width=0.78\linewidth]{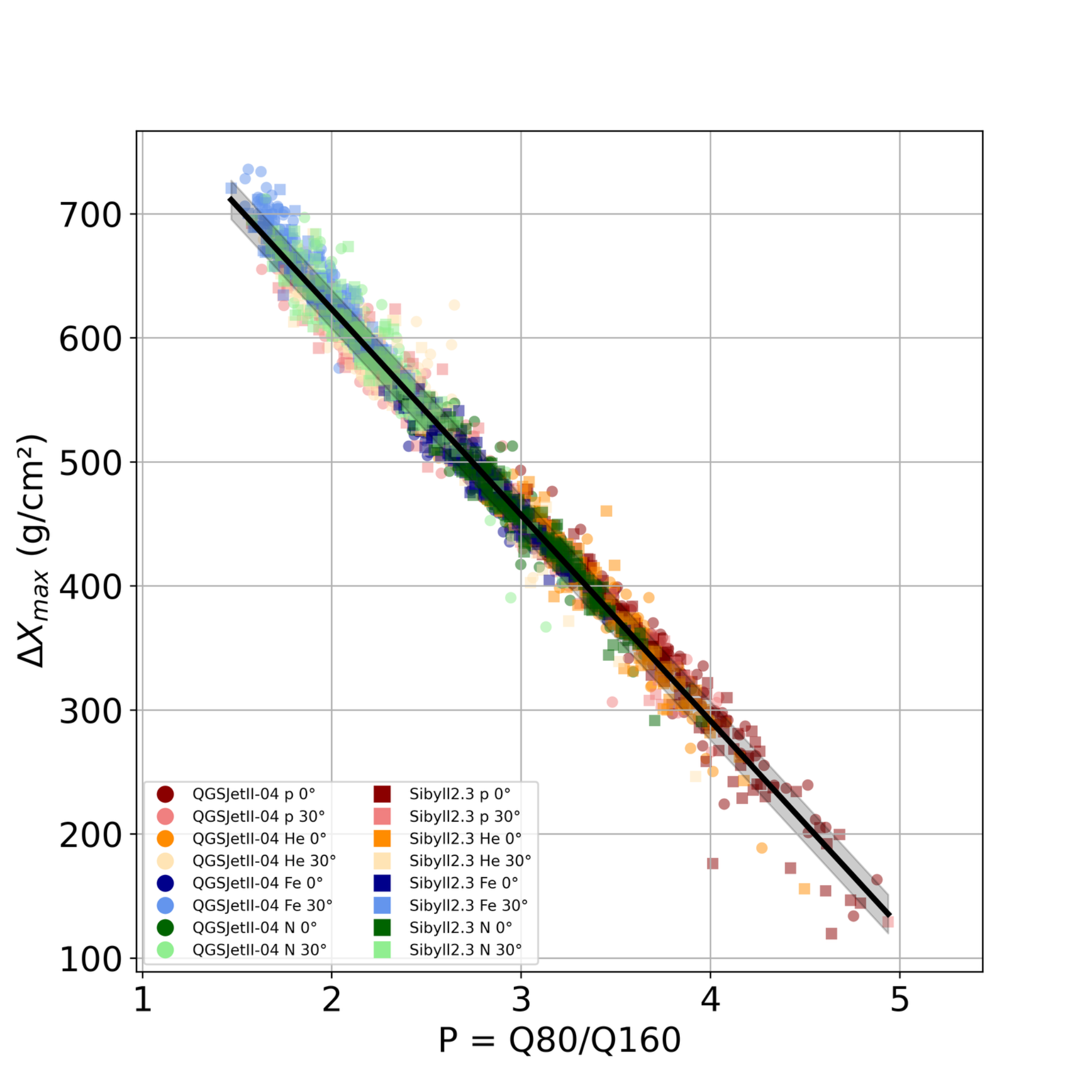}
\caption{Calculated dependence of $\Delta X_{\max}$ on the Cherenkov
light LDF steepness parameter $P=Q(80)/Q(160)$.}
\label{fig:p}
\end{figure}

Equation~\eqref{eq:pcalibration} is independent of primary-particle
energy over the range 1--30~PeV, zenith angle ($<30^\circ$), and
hadronic interaction model (QGSJET-II-04 or Sibyll~2.3d). The
simulated relationship is shown in Fig.~\ref{fig:p}. At energies above 30 PeV, this method is limited by saturation of the measured Cherenkov pulse amplitude at distances less than 100 m from the EAS axis.

\section{Analysis of experimental results}\label{sec:analysis}

\begin{figure}[H]
\centering
\includegraphics[width=0.85\linewidth]{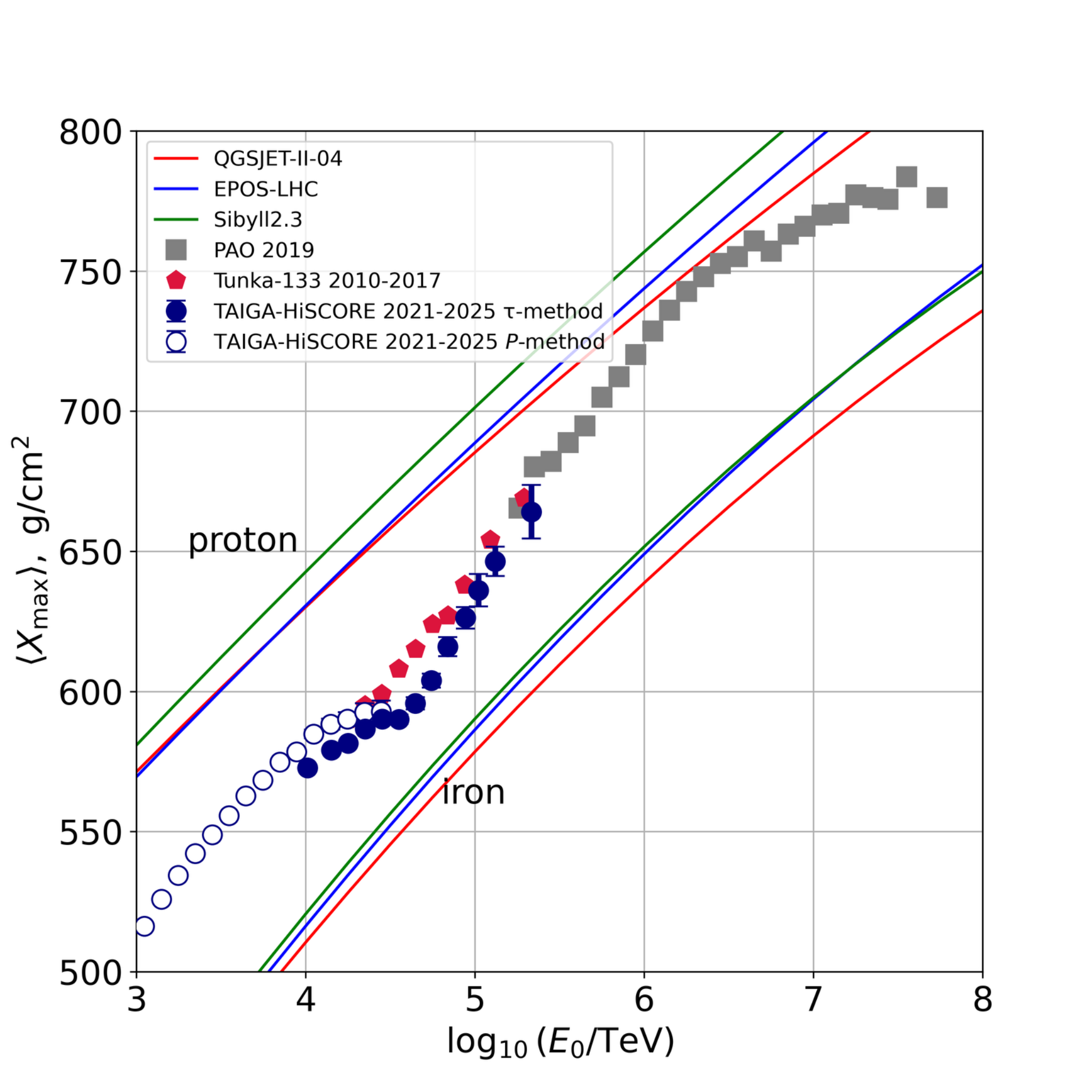}
\caption{Experimental dependence of the mean $\langle X_{\max}\rangle$ on the primary energy.}
\label{fig:xmax}
\end{figure}

The dependence of the mean shower-maximum depth,
$\langle X_{\max}\rangle$, on primary-particle energy was determined
from four TAIGA-HiSCORE observation seasons in 2021--2025 (about
900~hours of observation). Events were selected with zenith angles
$\theta\leq 30^\circ$ and shower cores within the effective detection
area of approximately 1~km$^2$. The energy range was 10--250~PeV for
the $\tau$-method (35,453 events) and 1--30~PeV for the $P$-method
(820,854 events). The experimental $\langle X_{\max}\rangle$ values
obtained with TAIGA-HiSCORE, together with results from other
experiments, are shown in Fig.~\ref{fig:xmax}.

The mean shower-maximum depth is converted into the mean logarithmic
mass $\langle\ln A\rangle$ using linear interpolation:

\begin{equation}
\langle\ln A\rangle =
\frac{\langle X_{\max}^{p}\rangle-\langle X_{\max}^{\mathrm{data}}\rangle}
{\langle X_{\max}^{p}\rangle-\langle X_{\max}^{\mathrm{Fe}}\rangle}
\,\ln 56.
\label{eq:lna}
\end{equation}

This interpolation was carried out using the QGSJET-II-04 model
\citep{ostapchenko2011}. The numerical $\langle X_{\max}\rangle$ and
$\langle\ln A\rangle$ values in the energy bins are listed in
Tables~\ref{tab:p} and~\ref{tab:tau}, respectively for the $P$- and
$\tau$-methods.

\begin{table}[htbp]
\centering
\normalsize
\caption{The $P$-method: TAIGA-HiSCORE data in the 1--30 PeV interval.}
\label{tab:p}
\begin{tabular}{@{}rrrrr@{}}
\toprule
$\lg(E/\mathrm{TeV})$ & $\langle X_{\max}\rangle$ (g/cm$^2$) & $\delta X_{\max}$ (g/cm$^2$) & $\langle\ln A\rangle$ & $\delta\langle\ln A\rangle$ \\
\midrule
3.045597 & 516.32 & 0.18 & 1.715 & 0.005 \\
3.147013 & 525.98 & 0.22 & 1.633 & 0.007 \\
3.246653 & 534.41 & 0.27 & 1.580 & 0.008 \\
3.346587 & 542.24 & 0.33 & 1.544 & 0.010 \\
3.446696 & 548.94 & 0.41 & 1.540 & 0.013 \\
3.546268 & 555.80 & 0.50 & 1.529 & 0.016 \\
3.646147 & 562.79 & 0.62 & 1.512 & 0.020 \\
3.746112 & 568.38 & 0.78 & 1.539 & 0.026 \\
3.845842 & 574.92 & 0.96 & 1.533 & 0.032 \\
3.945507 & 578.51 & 1.22 & 1.623 & 0.041 \\
4.045509 & 584.91 & 1.56 & 1.620 & 0.053 \\
4.146435 & 588.41 & 1.96 & 1.716 & 0.067 \\
4.246363 & 590.18 & 2.47 & 1.870 & 0.086 \\
4.346180 & 592.63 & 3.14 & 2.003 & 0.110 \\
4.445805 & 592.95 & 3.88 & 2.211 & 0.138 \\
\bottomrule
\end{tabular}
\end{table}

\begin{table}[htbp]
\centering
\normalsize
\caption{The $\tau$-method: TAIGA-HiSCORE data for energies from 10 PeV.}
\label{tab:tau}
\begin{tabular}{@{}rrrrr@{}}
\toprule
$\lg(E/\mathrm{TeV})$ & $\langle X_{\max}\rangle$ (g/cm$^2$) & $\delta X_{\max}$ (g/cm$^2$) & $\langle\ln A\rangle$ & $\delta\langle\ln A\rangle$ \\
\midrule
4.010847 & 572.78 & 0.44 & 1.955 & 0.015 \\
4.151481 & 579.11 & 0.76 & 2.046 & 0.026 \\
4.250379 & 581.53 & 0.92 & 2.179 & 0.032 \\
4.351593 & 586.73 & 1.17 & 2.222 & 0.041 \\
4.451388 & 590.17 & 1.41 & 2.322 & 0.050 \\
4.550986 & 590.13 & 1.62 & 2.548 & 0.058 \\
4.647211 & 595.84 & 2.17 & 2.559 & 0.079 \\
4.741832 & 603.93 & 2.46 & 2.478 & 0.090 \\
4.838065 & 616.06 & 3.40 & 2.247 & 0.126 \\
4.943829 & 626.33 & 3.81 & 2.100 & 0.143 \\
5.020429 & 636.15 & 5.81 & 1.900 & 0.219 \\
5.118750 & 646.51 & 5.23 & 1.723 & 0.199 \\
5.333590 & 664.14 & 9.57 & 1.511 & 0.373 \\
\bottomrule
\end{tabular}
\end{table}

\begin{figure}[H]
\centering
\includegraphics[width=0.85\linewidth]{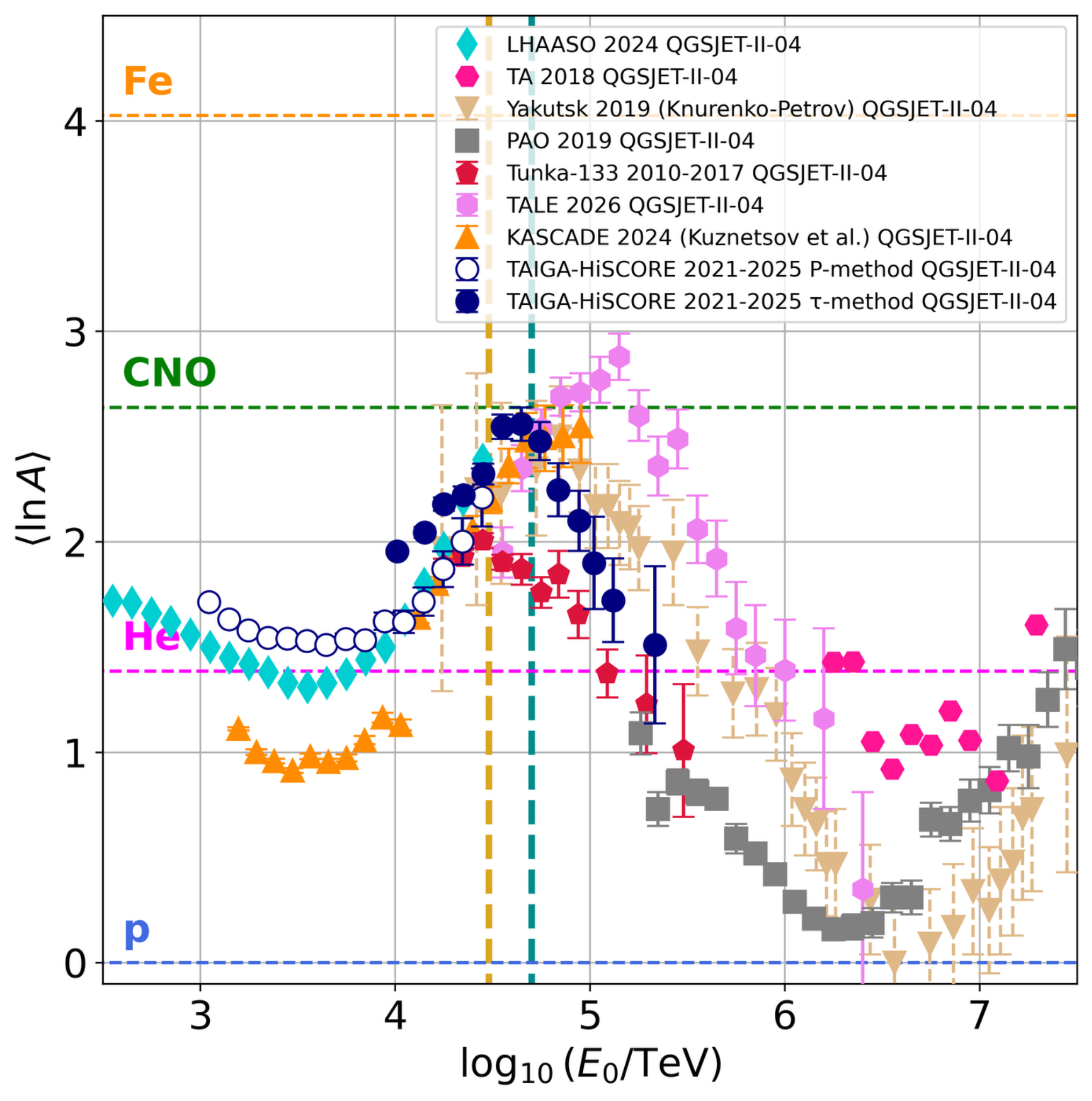}
\caption{Experimental dependence of the mean $\langle\ln A\rangle$ on the primary energy.}
\label{fig:lna}
\end{figure}

The conversion results are shown in Fig.~\ref{fig:lna}. According to
the TAIGA-HiSCORE measurements, the primary cosmic-ray composition
becomes heavier between the knee region and 30--50~PeV, approaching
the CNO-group value ($\langle\ln A\rangle\simeq2.6$), and then
becomes lighter at higher energies, reaching
$\langle\ln A\rangle\simeq1.75$ at about 300~PeV. The two
reconstruction methods yield shower-maximum depths differing by
approximately 10~g/cm$^2$ in this energy region. For comparison,
Fig.~\ref{fig:lna} includes results from LHAASO \citep{lv2025}, the
Pierre Auger Observatory \citep{yushkov2019}, Telescope Array
\citep{abbasi2018}, TALE \citep{abbasi2026}, KASCADE
\citep{kuznetsov2024}, the Yakutsk EAS array \citep{knurenko2019},
and Tunka-133 \citep{budnev2020}.

The mean mass compositions derived from the charged-particle component
of EAS by LHAASO (2024) and from the recently reanalysed KASCADE data
by Kuznetsov and colleagues (2024) exhibit a similar energy dependence
of $\langle\ln A\rangle$ to that obtained here.

Above 50--100~PeV, the TAIGA-HiSCORE points connect with those of the
Pierre Auger Observatory, which indicate a continuing shift toward a
lighter composition, approaching a purely protonic composition at
1,000~PeV. A trend toward lighter composition with increasing energy
is seen in the data of the large arrays considered here.

The conversion from shower-maximum depth to mean cosmic-ray mass
composition depends on the hadronic interaction model used in the EAS
simulations. The present analysis primarily uses QGSJET-II-04. Models
that predict more rapid dissipation of primary-particle energy at high
energies place the shower maximum higher in the atmosphere. Consequently,
for a primary-particle energy of 50~PeV and the same experimentally
measured depth of shower maximum, the inferred $\langle\ln A\rangle$
increases by approximately 0.35 (22.67\%) when using Sibyll~2.3d and by 0.08 (4.66\%) when using EPOS-LHC, relative to QGSJET-II-04.

\section{Conclusions}\label{sec:conclusions}

The methods for determining the EAS shower-maximum depth from
TAIGA-HiSCORE Cherenkov-array measurements have been improved: the
$\tau$-method reconstruction algorithm was refined and a new
$\Delta X_{\max}(P)$ calibration was obtained for the $P$-method.
Processing the 2021--2025 data with both methods yielded preliminary
values of $\langle X_{\max}\rangle$ and $\langle\ln A\rangle$ in
the energy range 1--250~PeV, which were compared with other
experimental results. The data support a shift toward a heavier
cosmic-ray mass composition from the knee to approximately 30~PeV,
followed by a shift toward a lighter composition above approximately
50~PeV. The difference between the results obtained by the two
$X_{\max}$-reconstruction methods is being investigated through
additional detailed simulations that account for the array geometry
and detector saturation close to the EAS axis. The current 
TAIGA-HiSCORE results should therefore be considered preliminary 
and subject to further revision.

\section*{Funding}
The work was carried out at the MSU-ISU
Astrophysical Complex (agreement EB 075-15-2021-675). The work was
supported by the Ministry of Science and Higher Education of the Russian
Federation (projects FZZE-2026-0006, FZZE-2027-0007, FSUS 2025-0008, FZZE-2025-0001) and
the Russian Science Foundation (grant 23-72-00016 (sections 3, 4, 5)).

\section*{Conflict of interest}
The authors declare no conflict of
interest.

\bibliographystyle{unsrtnat}
\bibliography{references}
\end{document}